\documentclass[twocolumn,amsmath,amssymb]{revtex4}
\usepackage{graphicx}
\usepackage{dcolumn}
\usepackage{bm}
\begin{document}

\title{The Pfaffian state in an electron gas with small Landau level gaps}
\author{Wenchen Luo and Tapash Chakraborty\thanks{
Tapash.Chakraborty@umanitoba.ca}}
\affiliation{Department of Physics and Astronomy, University of Manitoba, Winnipeg,
Canada R3T 2N2}
\date{\today}

\begin{abstract}
Landau level mixing plays an important role in the even denominator fractional
quantum Hall states. In ZnO the Landau level gap is essentially an order of
magnitude smaller than that in a GaAs quantum well. We introduce the screened
Coulomb interaction in a single Landau level to deal with that situation. Here
we study the incompressibility and the overlap of the ground state and the
Pfaffian (or anti-Pfaffian) state at filling factors $\frac52$ and $\frac72$
with a general screened Coulomb interaction. For small Landau level gaps,
the overlap is strongly system size-dependent and the screening can stabilize
the incompressibility of the ground state with particle-hole symmetry which
suggests a newly proposed particle-hole symmetry Pfaffian ground state. When
the ratio of Coulomb interaction to the Landau level gap $\kappa $ varies, we
find a possible topological phase transition in the range $2<\kappa <3$, which
was actually observed in an experiment. We then study how the width of quantum
well combined with screening influences the system.
\end{abstract}

\maketitle


The even-denominator fractional quantum Hall effect (FQHE) was observed
\cite{eisenstein,willett} and studied in great detail in GaAs. It is believed
that the concept of pairing of electrons is behind this unique quantum Hall
state \cite{Read,greiter,read2,son,willet2}, though the nature of this state
is still unclear. A strong candidate for the ground state is the Moore-Read
Pfaffian state which contains non-abelian excitations and chiral edge modes
\cite{Read,greiter,willet2}. However this topological state has not been
observed directly, perhaps because the mobility of GaAs is still not high
enough for this state to be detected. In other systems, such as cold atoms,
the circuit and cavity QED systems \cite{bloch,zhang,hafezi,hayward},
in theory it is possible to emulate this unique topological ground state by
tuning the Hamiltonian to approximate the parent Hamiltonian of the Pfaffian state.
The even-denominator FQHE has been studied and observed in graphene systems
\cite{Bilayer1,Bilayer2,papic,ki,abergeletal, Monolayer, FQHEGraphene,FQHEBilayer,julia,graphenebook}.
Recently, FQHE was observed again in the ZnO/MgZnO heterointerface
\cite{zno,tsukazaki,falson}. The
Pfaffian states and its topological properties can be potentially observed in
this new system, albeit
its low mobility. Surprisingly, in ZnO the well-known $\nu=\frac52$ FQHE was
found to go missing while its spinful electron-hole conjugate $\nu=\frac72$
FQHE survived \cite{falson,luo}. Tilted-field studies
\cite{tilted} also unveiled some interesting results in this system \cite{falson,luo2}.

The ZnO system is distinctly different from the GaAs and graphene systems
\cite{abergeletal,Monolayer,FQHEGraphene,FQHEBilayer,graphenebook,julia}
since the effective mass in ZnO is very large and the Landau level (LL) gap is very small.
The ratio of Coulomb interaction to the LL gap is $\kappa =25.1/\sqrt{B}$ for the magnetic
field $B$, which is one order of magnitude higher than that in GaAs or in graphene systems.
As a result, the electron-electron interaction would definitely drive the transport of the
electrons unconventionally in many aspects \cite{tapash}. Since the LL gaps are very small
in ZnO, the Landau level mixing (LLM) is too strong to be negligible. However, it would be
a major computational challenge to include even only a few LLs. In graphene or other small
$\kappa $ systems, perturbative theories which involve renomalized two-body and three-body
interaction have been developed in more accurate calculations \cite{perturbative}. However,
the limitation of that approach is that $\kappa $ cannot be too large. In experiments,
$\kappa $ is $0.5\sim 0.8$ (it depends on the dielectric constant of the substrate) for
graphene and could be smaller than unity (in a high magnetic field) for GaAs.
But $\kappa$ can not be smaller than unity in ZnO unless the magnetic field reaches 630T.
Consequently, we have proposed that
the screened Coulomb interaction in which all the other LLs are integrated out
in the random phase approximation (RPA) replaces the bare one, so that the correlations of
the electrons become very different at different filling factors. This seems to explain
the extraordinary phenomenon observed in the experiment \cite{luo,luo2}.

Trial wavefunctions have been proposed to describe this special even-denominator FQHE state
\cite{Read,halperin,hr,rr}. It appears that the spin polarized
Pfaffian state \cite{morf,spinpolarized} is the most probable candidate. Its particle-hole (PH) conjugate,
the anti-Pfaffian state \cite{antipf} is also likely since the two-body Coulomb interaction can
not break the PH symmetry in a half-filled Landau level. Recently, a new Pfaffian-like
state with PH symmetry \cite{son,zucker} was proposed and may be valid for
strong LLM or disorder. It is convenient to employ the Haldane
pseudopotential to study the overlap between the Pfaffian state and
the ground state of an even-electron system in the spherical geometry
\cite{haldane2,haldane3}. The rotational symmetry is preserved
so that the ground state of the incompressible state is uniquely located at
the total angular momentum $L=0$. In contrast, the ground state is degenerate
in the toroidal geometry due to the translational symmetry. In particular,
the ground state is quasi-triply degenerate for the even-electron system at
$\frac{5}{2}$ FQHE in toroidal geometry, due to the topological property
of the Pfaffian state \cite{peterson2}.

When the ratio $\kappa $ is very large the Coulomb interaction may not be
renormalized by the pertubation theory. Further, the PH symmetry may be
stabilized by the LLM \cite{son,zucker}. It means that the two-body screened
interaction in the RPA, which preserves the PH symmetry of the Hamiltonian, should
have essentially included some of the most important information of the LLM especially in
strong LLM, although it does not include all the correlations. We have
previously demonstrated in the torus geometry that the collective modes are
not stable for the $\nu=\frac52$ FQHE in ZnO in an odd-electron system.
However, in our previous works the nature of the ground states and the relation between
the screening and incompressbility still remained to be understood. In this work
the ground states are studied at different flux in a spherical geometry to
indicate how the ground states evolve with different screening.

The Pfaffian state is the zero-energy ground state of its parant Hamiltonian with a three-body
interaction \cite{greiter}. The pseudopotentials could be approximated instead for the planar case
\[
V_{m}=\int \frac{dq}{2\pi \ell}q \frac{V\left( q\right)}{\epsilon_s(q)} \left[ L^{}_n\left(
\frac{q^2}2\right) \right]^2L^{}_m\left( q^2\right) e^{-q^2},
\]
where $n$ is the LL index, $m$ is the momentum index, $\ell =\sqrt{\hbar /eB}$
is the magnetic length, and we add the screening effect into the system. The
screened pseudopotential was also used (although a simplified version of it)
in \cite{papic}. As in the torus geometry, the two-dimensional Coulomb
potential $V\left( q\right) =\frac{2\pi e^2}{\epsilon q}$ must be screened
by all the other LLs with the dielectric constant $\epsilon $. The static
dielectric function is $\epsilon_s^{}\left( q\right) =1-V(q) \chi^0_{nn}(q)$
\cite{luo,shizuya,luo3},
\begin{eqnarray*}
&&\chi^0_{nn}(q) =\frac{1}{2\pi \ell^2}
\sum_{\sigma,n,n^{\prime}}\frac{\min\left( n,n^{\prime }\right) !}{\max
\left( n,n^{\prime }\right) !}\left( \frac{q^2\ell^2}2\right)
^{\left\vert n-n^{\prime }\right\vert } \\
&&\times e^{-q^{2}\ell^2/2}\left[L_{\min \left( n,n^{\prime }\right)
}^{\left\vert n-n^{\prime }\right\vert }\left( \frac{q^2\ell^2}2
\right) \right]^2\frac{\nu_{\sigma ,n}^{}-\nu_{\sigma ,n^{\prime }}}{E_n^{}-E_{n^{\prime }}^{}},
\end{eqnarray*}
where $\sigma $ is the spin index, $E_{n}^{}$ is the kinetic energy of LL $n$,
$L(x)$ is the Laguerre function, and $\nu_{\sigma,n}^{}$ is the filling
factor. For simplicity, we consider only the non-interacting response
function. The dielectric function also preseves the rotational symmetry.

This planar screening with the kinetic energies are obtained by
the planar (not the spherical) LL quantization. In our numerical calculations
we use the planar $\epsilon_s$
approximately. The difficulty of the screening on a sphere is that the
degeneracies of different LLs are not the same at a constant flux, while the
number of the Landau levels are very limited and different LLs locate on spheres with different
radius if we consider a fixed degeneracy. This approximation would be more
accurate for a large sphere.

The flux on a sphere is $N_S=2N_{e}-S$ where $S$ is the shift with
a topological number \cite{wen}. The Pfaffian state occurs at $N_{Pf}=2N_{e}-3$,
while the anti-Pfaffian state occurs at $N_{aPf}=2N_{e}+1$. If we only
consider the two-body Coulomb interaction (the overlap of) the ground state of
$N_{e}$ electrons for $N_{Pf}$ is identical to that of $N_{e}-2$ electrons
for $N_{aPf}$.

\begin{figure}[tbp]
\includegraphics[width=7.0cm]{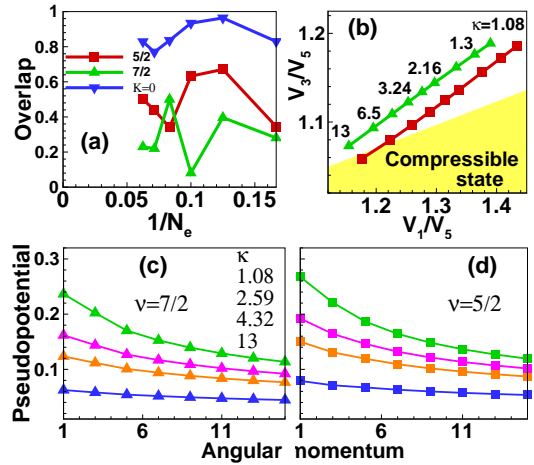}
\caption{(Color online) (a) The Pfaffian overlap at $\protect\nu =5/2$ and $
7/2$ with screening in ZnO, comparing with the unscreened case $\kappa=0$. 
(b)The trajectories of different
screened pseudopotentials at $\frac{5}{2}$ and $\frac{7}{2}$ fillings in the
pseudopoential $(V_1/V_5)-(V_3/V_5)$ plane with $\kappa \in [1.08, 13]$. The
compressible area is an extrapolation of the Ref. \cite{storni}. (c) and (d)
are the pseudopotentials for different screening.}
\label{figure1}
\end{figure}

In our previous work, we have shown that the spin is polarized when the two
LLs are very close or even crossed. Hence, only one spinless LL 
is considered here. We evaluated the collective modes and the ground states
for even number of electrons ($N^{}_e=4$ to $16$). For the ZnO system, we 
choose the experimental data,
$B=3.75$T at $\nu =\frac 52$ and $B=2.7$T at $\nu =\frac72$.

\begin{figure}[tbp]
\includegraphics[width=7.0cm]{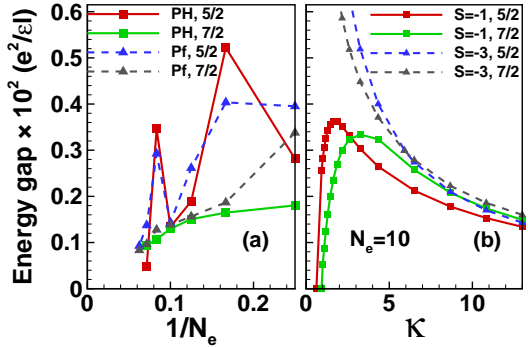}
\caption{(Color online) (a) The lowest energy gaps of the collective modes for
flux $N_{Pf}$ and $N_{PH}$ at fillings $\frac52$ and $\frac72$ in ZnO. (b) Energy
gaps vs. $\kappa$ for different flux with $10$ electrons.}
\label{figure2}
\end{figure}

The Pfaffian overlaps are shown in Fig.~\ref{figure1}(a), and the pseudopotentials
for different screening are shown in Figs.~\ref{figure1}(b) to (d). For
$\nu =\frac{5}{2}$, the overlap is decreased by half when $N^{}_e$ increases
to $12$. In contrast, for $\nu =\frac72$ the Pfaffian overlap dramatically
increases at $N^{}_e=12$. But when the system size increases
for up to 16 electrons the overlap of $\frac72$ decreases. From our
numerical works the overlap appears to be very size-dependent (much more than for the
unscreened case). On the other hand, the $(V_1/V_5)-(V_3/V_5)$ curve for
$\nu =\frac52$ is also very different from $\nu =\frac72$. In the strong LLM
region the ground state for $\nu =\frac52$ even enters into the compressible
area extrapolated from Ref. \cite{storni}, although we
do not find any compressibility for up to $16$ electrons. However, the ground
state should be compressible in the thermodynamic limit since the energy gap
in this case is very size dependent and oscillates very much in Fig.
\ref{figure2}(a), which agrees with our previous works in the less
size-dependent torus geometry \cite{luo}.

More importantly, in all cases the overlaps of the Pfaffian trial
wave function are not very high, but at most 0.65. This could be due to our
choice of planar screened dielectric function. It is also
possible that the ground state itself is another Pfaffian-like wave function
with PH symmetry \cite{son,zucker}, especially for large $\kappa$,
the strong LLM region. In the PH symmetric case where the shift is $S=-1$
and the flux is $N_{PH}=2N^{}_e-1$ on a sphere, the ground state is
compressible and located at $L=2$ for $N_e=10$ without screening and disorder.
So this PH symmetric flux was not considered in decades.

To confirm the possibility of the PH symmetric ground state in LLM, we
consider the PH symmetric flux. We calculate the energy spectrum for $N_e=10$
at this PH symmetric flux. In contrast, the ground states are surprisingly
incompressible for $\kappa>0.6$ at $\nu=\frac52$ and for $\kappa>0.93$ at
$\nu=\frac72$ when the screening is included in Fig. \ref{figure2}(b). More
generally, for $N_e =4 \sim 14$, all the ground states are stable and
incompressible with screening. Comparing with \cite{storni},
all the pseudopotentials $V_m$ here are modified by screening. $V_m$
with higher angular momentum also plays very important roles, since both the
overlap and the excitation gap are completely changed when $V_{m>5}=0$.
Moreover, if we only tune $V_{1,3}$ slightly and leave others unchanged,
the ground state still favors the Pfaffian shift $S=-3$. Only when all the
pseudpotentials are screened, then the particle-hole symmetric ground state
can be stabilized and incompressible. Hence, we assert here that screening
helps to stabilize the incompressibility of the PH symmetric states.

When the screening is weak (small $\kappa$), the excitation gaps at the
Pfaffian flux $N_{Pf}$ are a few times larger than those at PH flux $N_{PH}$
in Fig. \ref{figure2}(b).
So the Pfaffian state should be more stable. In the ZnO system ($\kappa
\approx 13\sim15 $) for both flux $N_{Pf}$ and $N_{PH}$, the lowest energy
gaps of $\frac{5}{2}$ are strongly size-dependent, while those gaps for
$\frac{7}{2}$ vary very smoothly as shown in Fig. \ref{figure2}(a). That is a
very strong indication that the $\frac52$ state is compressible in the
thermodynamic limit.
It seems that the $\frac72$ FQHE state would be closer to this newly
proposed Pfaffian-like state since the screening (the strength of LLM)
of $\nu=\frac72$ is always stronger than that of $\nu = \frac52$. Hence, the
overlap of the PH symmetry breaking Pfaffian or anti-Pfaffian
state is not high. On the other hand, our screened Coulomb interaction does 
not break the PH symmetry in a single LL.

\begin{figure}[tbp]
\includegraphics[width=7.0cm]{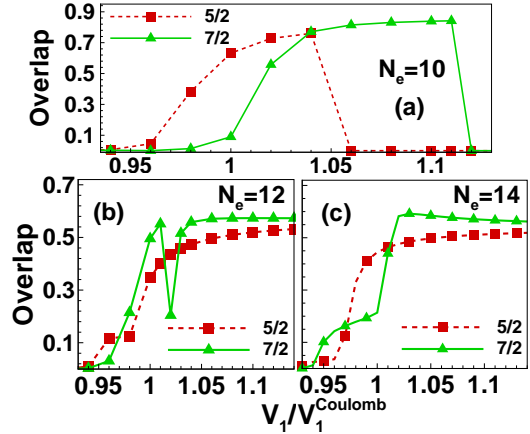}
\caption{(Color online) The overlap changes with the change of
pseudopotential $V^{}_1$ at (a) $10$, (b) $12$ and (c) $14$ electrons.}
\label{figure3}
\end{figure}

We artificially tune the pseudopotential $V^{}_1$ to study how the
pseudopotential influences the overlap of the topological
state in Fig.~\ref{figure3}. For $N^{}_e=10$, when $V^{}_1$ is varied
the $\frac72$ state would have much higher overlap than that of $\frac52$, and the
high-overlap window is much wider than that of $\frac52$. For
$N^{}_e=12$, the overlap of $\frac72$ state is generally higher than that of the
$\frac52$ state. For $N^{}_e=14$, the overlap of $\frac72$ gets steep rise
when $V^{}_1/V_1^{Coulomb}\approx1.01$. It is interesting that
the pseudopotential studies also provide indications that the two states
would be distinguished at different filling factors.

\begin{figure}[tbp]
\includegraphics[width=7.0cm]{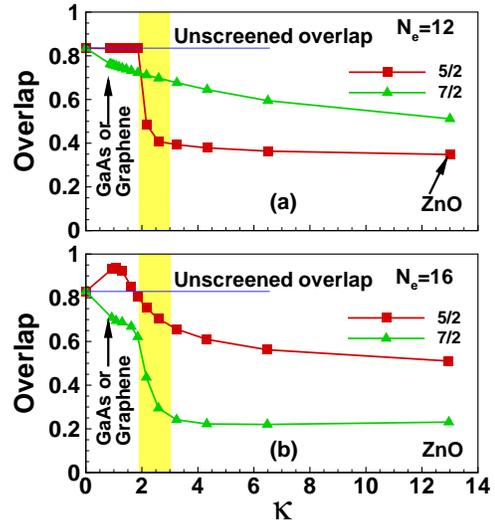}
\caption{(Color online) (a) The overlaps for $N^{}_e=12$ at different filling
factors are different. It suggests a phase transition between $\kappa =2$
and 3 for $\nu =5/2$. (b) For $N_{e}=16$, the overlap
of $\nu =5/2$ increase up to 0.94 when the screening is weak.  The rapid
drop occurs at $\nu =7/2$, also in the region $2< \kappa <3$.}
\label{figure4}
\end{figure}

We also calculate the Pfaffian overlaps with different screening since
screening is closely related to the LLM. Here we suppose the two filling 
factors are all at $B=3.75$T and $\kappa=13$. However, the results are also 
size-dependent, as shown in Fig.~\ref{figure4}. For $N_{e}=12$, in the ZnO
region the overlap of $\nu =7/2$ is higher than that of $\nu
=5/2$. The overlap of $\nu=7/2$ decreases in general, while at $\nu
=5/2$ for small $\kappa$ the overlap increases a little with screening and
falls rapidly in the region of $\kappa \in \lbrack 2,3]$. It
is possible that there is a topological phase transition between $\kappa =2$
and $3$. It seems that the numerical results are compatible with the
experiment in a doped GaAs system \cite{doppedGaAs}, where the
energy gap decreases dramatically at $\kappa \approx 2.6$, and decreases close
to zero at $\kappa \approx 2.9$. The decreased energy gap is the signal of
instability of the system. Results in Fig.~\ref{figure4}(a)
indicate that in the same region $\kappa \in \lbrack 2,3]$, the relation
between the ground state and the topological incompressible Pfaffian state
becomes gradually weaker. In Fig.~\ref{figure4}(b), for $N_{e}=16$ the
overlaps increase significantly up to $0.94$, when the screening is not
strong ($\kappa =1.08$). It decreases when the screening is strong. It
is interesting that this time the $\frac{7}{2}$ overlap dramatically drops
in the same region $\kappa \in \lbrack 2,3]$.

To confirm the stability of the ground states with different screening
strength ($\kappa $), we explore the scenario in torus
geometry \cite{luo,haldane,TCeven}. It is interesting
that when $\kappa >2.6,$ the ground state for $\nu =\frac{5}{2}$ becomes
unstable due to the softening of the collective modes. This shows the
geometry independence of the results. However, we do not find any
instability at $\nu =\frac{7}{2}$, though the gaps of the collective
modes are small. It is also possible that the $\frac{7}{2}$ state turns
into a PH Pfaffian-like state when the screening is increased, so that
the ground state is still incompressible.

All of these studies suggest that the ground state should be very close to
the Pfaffian state with large LL gaps. However, when the LLM
becomes stronger, i.e. $\kappa $ is increased, the ground state
could evolve into another state (perhaps the PH symmetric Pfaffian state)
and a topological phase transition may occur.
ZnO would be an ideal platform to experimentally study the PH symmetric Pfaffian
state at $\nu=\frac72$ due to the presence of strong LLM \cite{falson}.
However, more experiments, such as the thermal Hall conductance related to
the topological order of the bulk state \cite{antipf,zucker}, are necessary
to identify this property.

We also consider finite well thickness effect and consequently suppose that
the electron gas is trapped in an infinite square well with width $L_z$.
The $z-$component wave function is $\psi \left( z\right) =\sqrt{2/L_z}\sin
\left( n\pi z/L_z\right) $. We suppose that the well is not very wide so
that only the lowest subband dominates the system since the LL gap is very small
in ZnO. The Coulomb interaction is modified by multiplying a
thickness factor $V_{z}(q)$ \cite{peterson2}.
The screened dielectric function which still preserves the rotational
symmetry approximately becomes
\[
\epsilon'_{s}\left( q\right) =1-V_{z}\left( q\right) V\left( q\right)
\chi_{nn}^{0}\left( q\right).
\]


Thickness alone can not transform the
Pfaffian ground state to the PH symmetric one (when $L_z>5.8\ell$ it is possible,
but for a wide well the case should be different), yet with screening
the situation would change significantly.
For the Pfaffian shift $S=-3$, the $\frac{7}{2}$ overlap monotonically
increases with the increase of the width. But the overlap
of the $\frac52$ state decreases when the width begins to increase. Moreover,
the $\frac52$ overlap is also very size dependent, in contrast to the
monotonic increase in the absence of LLM \cite{peterson2}.
For the PH symmetric shift $S=-1$, the excitation gaps of $\frac72$ are not
sensitive to the width. The gaps of $\frac52$ become more smooth and less size
dependent when the width increases, which agrees with the results in the torus
geometry that the $\frac52$-FQHE is more stable in a wider ZnO quantum well
\cite{luo2}. Hence this agreement also supports the idea that
the ground state is more likely at the $-1$ shift for strong LLM.

To summarize, we have studied different topological states on a sphere
in a large region of $\kappa$. The screened Coulomb potential obtained
by the polarizability of all other LLs in the RPA indeed offers important
information about the LLM. It gives
us a clue to the nature of the even-denominator FQHE
with small LL gaps.
Since there is no direct
indication \cite{noevidence} of a phase transition between the Pfaffian
and anti-Pfaffian states in experiments, and further, the screening and
the thickness is
able to stabilize the incompressibility at the PH symmetric flux, it
reveals that a newly proposed PH symmetric Pfaffian state \cite{son,zucker}
may dominate the system in strong LLM. An obvious drop of the
Pfaffian overlap in the region of $\kappa \in \left[ 2,3\right]$
strongly suggests a topological phase transition from a Pfaffian state to
another.  If the strong LLM could be mapped onto the $s$-wave
scattering tunable \cite{chin} or fast rotating cold atom systems
(similar to the LLM effect) \cite{chang}, then the PH
symmetric Pfaffian state may be studied in those systems, which would open
up different venues for exploring new topological states of matter.

The work has been supported by the Canada Research Chairs Program of the
Government of Canada. The authors would like to thank V. Apalkov for helpful
discussions. The computation time was provided by Calcul Qu\'{e}bec and
Compute Canada.

\end{document}